\magnification=\magstep1
\voffset=-1.5truecm\hsize=16.5truecm    \vsize=24.truecm
\baselineskip=14pt plus0.1pt minus0.1pt \parindent=12pt

\lineskip=4pt\lineskiplimit=0.1pt      \parskip=0.1pt plus1pt 
\def\st{\scriptstyle} 
 
 
\input blackdvi   

 
\let\a=\alpha    \let\e=\varepsilon 
\let\f=\varphi        
      \let\p=\pi

\let\D=\Delta  \let\G=\Gamma

 
\global\newcount\numsec\global\newcount\numfor 
\gdef\profonditastruttura{\dp\strutbox} 
\def\senondefinito#1{\expandafter\ifx\csname#1\endcsname\relax} 
\def\SIA #1,#2,#3 {\senondefinito{#1#2} 
\expandafter\xdef\csname #1#2\endcsname{#3} \else 
\write16{???? il simbolo #2 e' gia' stato definito !!!!} \fi} 
\def\etichetta(#1){(\veroparagrafo.\veraformula) 
\SIA e,#1,(\veroparagrafo.\veraformula) 
 \global\advance\numfor by 1 
 \write16{ EQ \equ(#1) ha simbolo #1 }} 
\def\etichettaa(#1){(A\veroparagrafo.\veraformula) 
 \SIA e,#1,(A\veroparagrafo.\veraformula) 
 \global\advance\numfor by 1\write16{ EQ \equ(#1) ha simbolo #1 }} 
\def\BOZZA{\def\alato(##1){ 
 {\vtop to \profonditastruttura{\baselineskip 
 \profonditastruttura\vss 
 \rlap{\kern-\hsize\kern-1.2truecm{$\scriptstyle##1$}}}}}} 
\def\alato(#1){} 
\def\veroparagrafo{\number\numsec}\def\veraformula{\number\numfor} 
\def\Eq(#1){\eqno{\etichetta(#1)\alato(#1)}} 
\def\eq(#1){\etichetta(#1)\alato(#1)} 
\def\Eqa(#1){\eqno{\etichettaa(#1)\alato(#1)}} 
\def\eqa(#1){\etichettaa(#1)\alato(#1)} 
\def\equ(#1){\senondefinito{e#1}$\clubsuit$#1\else\csname e#1\endcsname\fi}

 
\def\bb{\hbox{\vrule height0.4pt width0.4pt depth0.pt}}\newdimen\u 
\def\pp #1 #2 {\rlap{\kern#1\u\raise#2\u\bb}} 
 
\def\ins #1 #2 #3 {\rlap{\kern#1\u\raise#2\u\hbox{$#3$}}}

\def\pallina{{\kern-0.4mm\raise-0.02cm\hbox{$\scriptscriptstyle\bullet$}}} 
\def\palla{{\kern-0.6mm\raise-0.04cm\hbox{$\scriptstyle\bullet$}}} 
\def\pallona{{\kern-0.7mm\raise-0.06cm\hbox{$\displaystyle\bullet$}}} 
\def\bull{\vrule height .9ex width .8ex depth -.1ex } 
 
\def\data{\number\day/\ifcase\month\or gennaio \or febbraio \or marzo \or 
aprile \or maggio \or giugno \or luglio \or agosto \or settembre 
\or ottobre \or novembre \or dicembre \fi/\number\year} 
 
\setbox200\hbox{$\scriptscriptstyle \data $} 
 
\newcount\pgn \pgn=1 
\def\foglio{\number\numsec:\number\pgn 
\global\advance\pgn by 1} 
\def\foglioa{a\number\numsec:\number\pgn 
\global\advance\pgn by 1} 
 
\footline={\rlap{\hbox{\copy200}\ $\st[\number\pageno]$}\hss\tenrm \foglio\hss} 
 
 
\def\sqr#1#2{{\vcenter{\vbox{\hrule height.#2pt 
\hbox{\vrule width.#2pt height#1pt \kern#1pt 
\vrule width.#2pt}\hrule height.#2pt}}}}

 \def\\{\noindent}

\let\dpr=\partial

 \def\AA{{\cal A}}

\def\tende#1{\vtop{\ialign{##\crcr\rightarrowfill\crcr 
              \noalign{\kern-1pt\nointerlineskip} 
              \hskip3.pt${\scriptstyle #1}$\hskip3.pt\crcr}}} 
\def\otto{{\kern-1.truept\leftarrow\kern-5.truept\to\kern-1.truept}} 
 
\font\smfnt=cmr10 scaled\magstep0 
\font\myfonta=msbm10 scaled \magstep0 
\def\math#1{\hbox{\myfonta #1}}


\vglue.5truecm 
{\centerline{\bf ON THE CONVERGENCE TO THE CONTINUUM OF FINITE RANGE}}  

{\centerline{\bf LATTICE COVARIANCES}}

\vglue1.truecm 
{\centerline{{\bf David C.Brydges}$^{1}$, {\bf P.K.Mitter}$^{2}$}} 

\vglue.5truecm 
{\centerline{\smfnt$^1$ Department of Mathematics, The University of British
Columbia}} 
{\centerline{\smfnt 1984 Mathematics Road, Vancouver, British Columbia,
Canada V6T1Z2}} 
{\centerline{\smfnt e-mail: db5d@math.ubc.ca}} 
 
\vglue.5truecm 
{\centerline{\smfnt$^2$ D\'epartement de Physique Th\'eorique,
Laboratoire Charles Coulomb}} 
{\centerline{CNRS-Universit\'e Montpellier 2}} 
{\centerline{\smfnt Place E. Bataillon, Case 070,  
34095 Montpellier Cedex 05 France}} 
{\centerline{\smfnt e-mail: pmitter@univ-montp2.fr}} 

\vglue1cm

\\{\bf Abstract}: In J Stat Phys. {\bf 115}, 415-449 (2004) 
Brydges, Guadagni and Mitter proved the existence of multiscale expansions
of a class of lattice Green's functions as sums of positive definite finite
range functions (called fluctuation covariances). The lattice Green's 
functions in the class considered are integral kernels of inverses of 
second order positive self-adjoint elliptic operators with constant
coefficients and fractional powers thereof. 
The fluctuation covariances satisfy uniform bounds and the sequence converges
in appropriate norms to a smooth, positive definite, finite range
continuum function.
In this note we prove that the convergence is actually exponentially fast. 

\vglue.5truecm
\\{\bf 1. Introduction} 
\numsec=1\numfor=1 
\vglue.5truecm

In [BGM], Brydges, Guadagni and Mitter proved the existence of
multiscale expansions of a class of lattice Green's functions as sums
of positive definite finite range functions (fluctuation
covariances). The lattice Green's functions that were considered are
integral kernels of inverses of self-adjoint lattice elliptic
operators. The construction in [BGM] was given $1$) for the resolvent
operator $(a-\D)^{-1}$ with $a\ge 0$ on $\math{Z}^{d}$ where $\D$ is
the standard lattice Laplacian, the resolvent parameter $a\ge 0$, and
$d\ge 3$ and $2$) for the L\'evy Green's function $(-\D)^{-\a/2}$ on
$\math{Z}^{d}$ with $d\ge 3$, and $0<\a <2$. This has been extended in
[BT] to Green's functions of more general self-adjoint elliptic
operators. The summands, called fluctuation covariances, after
rescaling live on finer and finer lattices, have uniformly bounded
support (finite range property) and it was proved in [BGM] that their
Fourier transforms satisfy bounds independent of lattice spacing and
have strong decay properties. It was also proved that the sequence of
rescaled fluctuation covariances converge in appropriate norms to a
smooth positive definite finite range continuum function. In the
present note (which is a sequel to the above paper and should be read
as such) we prove that the convergence is exponentially fast. This is
of some importance for renormalization group applications. An example
is furnished in the work of Mitter and Scoppola, [MS]. The exponential
convergence is stated in Theorem 1.1, page 931 of [MS] and then used
in the construction of the stable manifold in section 6 of that
paper. The present work furnishes the promised proof of that
result. Another application is in the forthcoming work of
R. Bauerschmidt (in preparation) where, amongst other things,
exponential convergence is extended to the mass derivative of the
finite range expansion.  \Red {Applications} of finite range
multiscale expansions in rigorous renormalization group analysis
include the work of Brydges and Slade on weakly self-avoiding simple
random walks in $\math{Z}^{d}$, $d\ge 4$ [BS], a new proof of the
thermodynamic limit for the dipole gas by Dimock [D], the
critical line in the Kosterlitz-Thouless transition by P. Falco [F]
and forthcoming work by Stefan Adams et al. on gradient models.

\vglue.5truecm
\\{\bf 2. Summary of earlier results and main theorem} 
\numsec=2\numfor=1 
\vglue.5truecm

In this section we will first summarize the results in [BGM] to the
extent we will need them in order to be able to state our basic
convergence estimates (Theorem~2.1, Corollaries~2.2 and 2.3).  The
proof of Theorem~2.1 will be given in section 3.  Throughout this
paper we will use the notations, definitions and results given in
[BGM]. Let $L=2^{p}$ be a dyadic integer. $L$ must be chosen
sufficiently large depending on the dimension $d$ and on the rate of
decay given by the parameter $k$ that appears in all our estimates
starting with (1.13).

It is assumed to be large We define $\e_n=L^{-n}$, $n\ge 0$. For
$n=0,1,2,...$ we have a sequence of lattices
$(\e_{n}\math{Z})^{d}\subset {\bf R}^{d}$ which are nested,
$(\e_{n}\math{Z})^{d}\subset (\e_{n+1}\math{Z})^{d}$.  We assume $d\ge
3$.

\vskip0.2cm

\vskip0.2cm

Let $\D_{\e_n}$ be the lattice Laplacian acting on functions on
$(\e_{n}\math{Z})^{d}$. For $a\ge 0$ the resolvent
$G^{a}_{\e_{n}}=(-\Delta_{\e_{n}} + a)^{-1}$ has the Fourier transform

$$G^{a}_{\e_{n}}(x-y)=\int_{B_{e_n}}
{d^{d}p\over{(2\pi)^{d}}}\>
e^{ip.(x-y)} (a-\hat{\D}_{\e_n}(p))^{-1} \Eq(1.1) $$

\\where $B_{e_n}=[{-\pi\over\e_{n}},{\pi\over\e_{n}}]^{d}$ and 

$$\hat{\D}_{\e_n}(p)=2\e_{n}^{-2}\sum_{\mu =1}^{d}\big(
\cos(\e_{n} p_{\mu})-1\big).\Eq(1.2) $$

\\Let $U_{c}(R_{m})=(-{R_{m}\over 2}, {R_{m}\over 2})^{d}\subset
(\math{R})^{d}$ denote a continuum cube of edge length
$R_m=L^{-(m-1)}$. Here $m=0,1,2...,n$. Then  $U_{\e_n}(R_m)=
U_{c}(R_{m})\cap (\e_{n}\math{Z})^{d} $ defines a cube in the lattice
$(\e_{n}\math{Z})^{d}$.
The boundary $\dpr U_{\e_n}(R_m)$ is defined to consist of lattice points 
not in $U_{\e_n}(R_m)$ which have a nearest neighbour in $U_{\e_n}(R_m)$.

{\it Remark}: The choice $L=2^{p}$, in particular that $L$ is even,
implies that the boundary of the continuum cube passes through lattice
points. Therefore the boundary $\dpr U_{\e_n}(R_m)$ of the lattice
cube is contained in the boundary $\partial U_{c}(R_{m})$ of the
continuum cube. This is used in the proof of Lemma 6.5 of [BGM]. If
one prefers, for example $L=3^{p}$, then replacing $R/2$ by $R/3$ in
the definition of $U_{c}(R_{m})$ retains this boundary property.

A measure on the lattice is just a weighted sum of point masses at
lattice points, but it facilitates comparison with the continuum to
write sums as integrals over such measures.  ${\cal P}^{a}_{\dpr
U_{\e_n}(R_m)}(x,du)$ denotes the Poisson kernel measure on $\dpr
U_{\e_n}(R_m)$.  By definition this is such that if $f$ is a function
on $\dpr U_{\e_n}(R_m)$ then

$$h_{\e_n,m}(x)= {\cal P}^{a}_{\dpr U_{\e_n}(R_m)}(x,f)
=\int_{\dpr U_{\e_n}(R_m)}{\cal P}^{a}_{\dpr U_{\e_n}(R_m)}(x,du)f(u)
\Eq(1.3)$$ 

\\solves the Dirichlet problem

$$(-\D_{\e_n}+1)h_{e_n,m}(x)=0:\quad x\in U_{\e_n}(R_m) \Eq(1.4) $$

$$h_{\e_n,m}(x)=f(x):\quad x\in \dpr U_{\e_n}(R_m).\Eq(1.5)$$

\\The Poisson kernel measure exists and a probabilistic representation was
given (and exploited) in [BGM]. For $a=0$, the  Poisson kernel measure is
a probability measure, otherwise ($a>0$) it is a defective measure 
(total mass  is less than $1$). In [BGM] an averaging map 
$f\rightarrow A^{a}_{\e_n,m}(R_m)f$ was introduced for functions $f$
defined on $(\e_n \math{Z})^{d}$. This uses the Poisson kernel measure above.
In the next paragraph we recall the definition of the averaging operation and 
then the fluctuation measures which enter in the finite range multiscale 
expansion of Green's functions established in [BGM].
   
Let $g$ be a non-negative, rotationally invariant, $C^{\infty}$  
function on $\math{R}^{d}$  of finite range 
$L\over 4$. In other words $g(x)$ vanishes for $|x|\ge {L\over 4}$. 
$g$ is chosen to be normalized so that $\int_{\math{R}^{d}} dx\> g(x)=1$. 
Define the sequence of functions $g_n$ by $g_n(x)= L^{nd}g(L^{n}x)$ for
$n=0,1,2...$. Then the functions $g_n$ have mass $1$ and finite range 
${1\over 4}L^{-(n-1)}$. Restrict $g$ to the 
lattice $(\e_n \math{Z})^{d}$ and let $c_{\e_n}$ be the positive constant so 
that  $c_{\e_n}\int_{(\e_n \math{Z})^{d}} dx\> g(x)=1$. Here the integration 
is with respect to the ``Lebesgue'' measure on the lattice $(\e_n \math{Z})^{d}$  
i.e. the counting measure times $\e^{d}_n$. The constants $c_{\e_n}$
converge to $1$ as $n\rightarrow \infty$.
We have
$\int_{(\e_n \math{Z})^{d}} dx\>c_{\e_{n-m}} g_m(x)
=\int_{(\e_{n-m} \math{Z})^{d}} dx\>c_{\e_{n-m}} g(x)= 1$. Let $f$ be a 
function on $(\e_n \math{Z})^{d}$. For $m=0,1,...n$ we define a sequence
of (averaging) maps  $f\rightarrow A^{a}_{\e_n,m}(R_m)f$ 
and their kernels $A^{a}_{\e_n,m}(R_m)(x,u)$ by 

$$(A^{a}_{\e_n,m}(R_m)f)(x)=
\int_{(\e_n \math{Z})^{d}} dz\>c_{\e_{n-m}} g_{m}(z-x)
{\cal P}^{a}_{\dpr U_{\e_n}(R_m)}(x-z,f(z+\cdot)) $$

$$=\int_{(\e_n \math{Z})^{d}}du\> A^{a}_{\e_n,m}(R_m)(x,u)f(u)\Eq(1.6)$$

\\where (see [BGM], page 423-424)  $A^{a}_{\e_n,m}(R_m)(x,u)\>du$ is a 
family of translation invariant 
(defective) probability measures
on $(\e_n \math{Z})^{d}$. The support 
property of $g_m$ makes sure that the Poisson kernel entering above is never 
evaluated on $x$ near the boundary point $u$ where derivatives become large. 

Consider first the case $m=0$ and recall that $R_0 =L$. We define a
fluctuation covariance  

$$\G^{a}_{\e_n}(x-y)=G^{a}_{\e_n}(x-y)-
(A^{a}_{\e_n,0}(R_0)G^{a}_{\e_n}A^{a}_{\e_n,0}(R_0)^{*})(x-y). \Eq(1.7)$$

\\$\G^{a}_{\e_n}$ is a positive definite function of finite range $L$ and
$\hat\G^{a}_{\e_n}(p)$ is continuous in $p$ including at $p=0$, (Lemma~3.1,
[BGM]). For $n\ge 1$ define

$$\G^{a}_{n}=\AA^{a}_{n} \G^{a}_{\e_n}{\AA^{a}_{n}}^{*}\Eq(1.8)$$
\\where

$$\AA^{a}_n =\prod_{m=1}^{n}A^{a}_{\e_n,m}(R_m)\Eq(1.9)$$

\\and the product above is given by a multiple convolution. For $n=0$
we set $\AA^{a}_0 =1$. $\G^{a}_{n}$ is a positive definite function with 
finite range $6L$, (Lemma~3.2, [BGM]). Let $G^{a}=: G^{a}_{\e_0}$ be the
unit lattice resolvent. 
$$G^{a}(x-y)=\sum_{n\ge 0} L^{-n(d-2)} \G^{a_n}_{n}({x-y\over L^{n}})\Eq(1.10)$$ 

\\where $a_n=L^{2n}a$.

\vglue0.2cm
\\{\it Remark 1}: The factor $6$ in the range $6L$ of $\G^{a}_{n}$ is an 
artifact. By scaling down the edge length $R_m=L^{-(m-1)}$ of the cube 
$U_{\e_n}(R_m)$ to $R_m={1\over 16}L^{-(m-1)}$ and the range of $g$ from
$L/4$ to $L/64$ we get $\G^{a}_{n}$ to have finite range (less than) $L/2$.

\vglue0.2cm
\\Let $G=(-\D)^{-{\a\over 2}}$, $0<\a<2$,
be the Green's function of a L\'evy walk in $\math{Z}^{d}$. $G$ has the 
integral representation $G={\rm const}\int_{0}^{\infty}da\> a^{-\a/2}G^{a}$.
Integrating \equ(1.10) with the measure $da\>a^{-\a/2}$ we get the 
finite-range multiscale expansion for $G$

$$G(x-y)=\sum_{n\ge 0} L^{-n[\f]} \G_{n}({x-y\over L^{n}})\Eq(1.11)$$

\\where $[\f]={d-\a\over 2}$ and 

$$\G_{n}=\int_{0}^{\infty}da\> a^{-\a/2}\G^{a}_{n}. \Eq(1.12)$$

\\These formulae make sense by virtue of the following bound provided in 
Theorem~5.5, page 434, [BGM]:

\vglue0.2cm
\\Let $B_{\e_n}=[{-\p\over \e_n},{\p\over \e_n}]^{d}$, the first
Brillouin zone of the dual lattice. Then for all $n\ge 0$ and all $
k\ge 0$, there is a constant $c_{k,L}$ independent of $n$ such that
for $p\in B_{\e_n}$,

$$|\hat\G^{a}_{n}(p)|\le c_{k,L}(1+a)^{-1}(1+p^2)^{-2k}.  \Eq(1.13)$$

\vglue0.2cm
\\{\it Remark 2}: For $n\ge 1$, the above bound can be improved to

$$|\hat\G^{a}_{n}(p)|\le c_{k,L}e^{-ca^{1\over 2}}(1+p^2)^{-2k} \Eq(1.14)$$ 

\\where the constant $c$ is a positive length which does not depend on 
$L$ or the indices $k,n$ but may depend  on the dimension. 
Since $a$ is replaced by $a_{n}$ in \equ({1.10}) 
the finite range multiscale expansion has double exponential convergence for $a$ 
positive!

{\it Proof:} This follows on exploiting the remark
on exponential decay on page 445, [BGM] in the proof of Proposition~5.2, page
432 and then following its consequences up to page 435. For $n\ge 1$,
this leads to the bound
on $\hat\AA^{a}_{n}(p)$ on page 435, [BGM] being improved to

$$|\hat\AA^{a}_{n}(p)|^{2}\le c_{k,L} e^{-ca^{1\over 2}}(1+p^2)^{-2k}. 
\Eq(1.15)$$ 

The $e^{-ca^{1\over 2}}=e^{-ca^{1\over 2}R_{1}} $ comes from
the $m=1$ factor of $A^{a}_{\e_n,m}(R_m)$ in
\equ({1.9}). Since  $R_1 =1$ the length coefficient of $a^{1\over 2}$ in the exponent
is $O(1)$. The Fourier transform of $\G^{a}_n$ is given by

$$\hat\G^{a}_{n}(p)=|\hat\AA^{a}_{n}(p)|^{2}\hat\G^{a}_{\e_n}(p). \Eq(1.16)$$

\\Combining \equ(1.15) with the bound 
$|\hat\G^{a}_{\e_n}(p)| \le c_L (1+p^2)^{-1} $ 
gives the bound \equ(1.14). \bull

As explained in Section~6 of [BGM] the construction of the finite range
multiscale expansion above can be directly done in exactly the same way
in $\math{R}^{d}$. The cube $U_{\e_n}(R_m)$ is now replaced 
by the continuum cube $U_{c}(R_m)$ where we follow here and hereafter
the convention that the subscript $c$ denotes that we are in the continuum. 
The lattice resolvent $G^{a}_{\e_{n}}$ is replaced by the continuum resolvent 
$G^{a}_{c}$ with Fourier transform $\hat G^{a}_{c}(p)=(a+p^2)^{-1}$. The
solution of the continuum analogue of the Dirichlet problem \equ(1.3)
-\equ(1.5) is denoted by
$h^{a}_{c,m}(x)= {\cal P}^{a}_{\dpr U_{c}(R_m)}(x,f)$ where 
${\cal P}^{a}_{\dpr U_{c}(R_m)}$ is the continuum Poisson kernel measure.
The continuum averaging
operation $f\rightarrow A^{a}_{c,m}(R_m)f$ is now defined as in \equ(1.6)
using the continuum Poisson kernel measure and the $z$-integration is 
in $\math{R}^{d}$. The continuum fluctuation covariances $\G^{a}_{c}$, and   
$\G^{a}_{c,n}$ are defined by the continuum analogues of \equ(1.7)-\equ(1.9).
Using Fourier transforms we have

$$\hat\G^{a}_{c}(p)=\hat G^{a}_{c}(p)- |\hat A^{a}_{c,0}(R_0)(p)|^{2}
\hat G^{a}_{c}(p)\Eq(1.17) $$

$$\hat\G^{a}_{c,n}(p)=|\hat\AA^{a}_{c,n}(p)|^{2}\hat\G^{a}_{c}(p)
=\prod_{m=1}^{n}|\hat A^{a}_{c,m}(R_m)(p)|^{2}\hat\G^{a}_{c}(p). \Eq(1.18) $$

\\The continuum analogue of the elliptic estimates ( Appendix A, [BGM])
used in the proof of Theorem~5.5 of [BGM]
imply that the bounds \equ(1.13),\equ(1.14) continue to hold in 
$\math{R}^{d}$ for $n\ge 1$. Thus we have that for all $n\ge 1$ and all
$k\ge 0$ there is  a constant $c_{k,L}$ independent of $n$ such that

$$|\hat\G^{a}_{c,n}(p)|\le c_{k,L}(1+a)^{-1}(1+p^2)^{-2k} \Eq(1.19)$$

\\and its improvement

$$|\hat\G^{a}_{c,n}(p)|\le c_{k,L}e^{-ca^{1\over 2}}(1+p^2)^{-2k}.
\Eq(1.20)$$ 

\\The following statements are proved in Section~6 of [BGM], (see Theorem~6.1
and its proof).

\vglue0.2cm

\\1. {\it Continuum covariances:} The uniformly bounded sequence $\{\hat\G^{a}_{c,n}(p)\}_{n\ge 1}$, (see
above), is Cauchy so that pointwise in $p$ ,

$$\hat\G^{a}_{c,n}(p) \rightarrow \hat\G^{a}_{c,*}(p) \Eq(1.21)$$

\\and $\hat\G^{a}_{c,*}(p)$ satisfies the bounds \equ(1.19), \equ(1.20).  
$\G^{a}_{c,*}(x)$ is in $H_{k}(\math{R}^{d})$ for all $k\ge 0$. Therefore
by Sobolev embedding $\G^{a}_{c,*}(x)$ is a smooth function.

\\2. {\it Lattice covariances:} Pick any integer $l\ge 1$ and let $p\in B_{\e_l}$. 
For $n\ge l$ the sequence $\{\hat\G^{a}_{n}(p)\}_{n\ge 1}$, (see
\equ(1.13), \equ(1.14)) converges to the continuum limit function above: 
 
$$\hat\G^{a}_{n}(p) \rightarrow \hat\G^{a}_{c,*}(p).\Eq(1.22)$$

\\The next theorem, which is our main result, shows that the
convergence is exponentially fast.  It is stated in terms of a Sobolev
space $L^{1}_{k}((\e_{l}\math{Z})^{d})$ which is discussed below the
theorem.

\vglue0.2cm

\\{\it Theorem~2.1}: Pick any integer $l\ge d$.  Restrict
$\G^{a}_{c,*}$ to $(\e_{l}\math{Z})^{d}$.  Then for all $n\ge l$ and
all $k\ge 0$ there is a constant $c_{k,L}$ independent of $n$ such
that

$$\Vert \G^{a}_{n}-\G^{a}_{c,*}\Vert_{L^{1}_{k}((\e_{l}\math{Z})^{d})}
\le c_{k,L}L^{-{n\over 2}} e^{-ca^{1\over 2}}. \Eq(1.23)$$

\\{\it Remark}: Let $\Omega \subset \math{R}^{d}$ be an open set. Let
$C_{0}^{\infty} (\Omega)$ be the space of $C^{\infty}$ functions of
compact support in $\Omega$. Then $L^{1}_{k}(\Omega)$ is the Banach
space (also known as $W^{1,k}_{0}(\Omega)$) 
obtained by completing $C_{0}^{\infty} (\Omega)$ in the norm

$$\Vert f\Vert_{L^{1}_{k}(\Omega)}=
\sum_{0\le |\a|\le k}\Vert D^{\a}f\Vert_{L^{1}(\Omega)}.\Eq(1.231) $$

\\Let now $\Omega$ be a bounded open cube. Then, as is well known,
repeated application of the Poincar\'e inequality
gives the equivalent norm (see e.g. [A])

$$\Vert f\Vert_{L^{1}_{k}(\Omega)}= \sum_{|\a|=k}\Vert D^{\a}f\Vert_{L^{1}(\Omega)}\Eq(1.232) $$

\\The same definitions are adapted to the lattice with integrals and
derivatives being replaced by sums and finite differences (forward
lattice derivatives). Just as in the continuum the equivalent norm is
proved by repeated applications of the lattice Poincar\'e inequality
(proved in Lemma~B2 of Appendix B of [BGM]).  The $L^{1}_{k}$ Sobolev
spaces of index larger than $d$ embed into spaces of continuous
functions (see e.g [A]) and the same proof works in the continuum and
the lattice. This can be seen in the proof of Lemma~B.1 of Appendix B
of [BGM] in which the first equation together with the argument in the
last four lines of the proof implies, for $k > d+j$, that

$$\Vert f\Vert_{C^{j}_{0}(\Omega)} \le C_{\Omega,j,k}
\Vert f\Vert_{L^{1}_{k}(\Omega)}\Eq(1.233) $$

\\where the lattice norm denoted by $C^{j}_{0}(\Omega)$ is defined as
the supremum over the lattice derivatives of orders up to $j$ of
functions of compact support in $\Omega$.  We can use the spaces
obtained by completing smooth functions of compact support because
$\G^{a}_{n}(x),\>\G^{a}_{c,*}(x)$ are of finite range $6L$, i.e. they
vanish for $|x|\ge 6L$. The norm in \equ(1.23) can therefore be taken
in the finite cube $\Omega_{\e_n}=U_{\e_n}(6L)$.

\vglue0.3cm

\\Let $\dpr^{\a}_{\e_n}=\prod_{j=1}^{d}\dpr_{\e_n,e_j}^{\a_{j}}$, 
$\a=(\a_1,...,\a_d)$, $\a_j$ non-negative integers, 
denote a multiple $\e_n$-lattice partial 
derivative. Here $\dpr_{\e_n,e_j}^{\a_{j}}$ is the forward
$\e_{n}$-lattice derivative in direction $e_j$. The $e_1,..,e_d$ are unit vectors 
specifying the orientation of $\math{R}^{d}$ and all embedded lattices 
$(\e_{n}\math{Z})^{d}$. Let $\dpr^{\a}_{c}$ be a multiple continuum
partial derivative. Then \equ(1.23) implies by Sobolev 
embedding of high degree lattice $L^{1}_{k}$ spaces: 

\vglue0.2cm
\\{\it Corollary~2.2}: For all $|\a|\ge 0$, and for all $n\ge l\ge d$    

$$\Vert \dpr^{\a}_{\e_n} \G^{a}_{n}-
\dpr^{\a}_{c}\G^{a}_{c,*}\Vert_{L^{\infty}((\e_{l}\math{Z})^{d})}
\le c_{k,L}L^{-{n\over 2}} e^{-ca^{1\over 2}}. \Eq(1.24)$$

\\{\it Proof}: We have
$$\Vert \dpr^{\a}_{\e_n} \G^{a}_{n}-
\dpr^{\a}_{c}\G^{a}_{c,*}\Vert_{L^{\infty}((\e_{l}\math{Z})^{d})}$$
$$
\le
\Vert\dpr^{\a}_{\e_n} \G^{a}_{n}-
\dpr^{\a}_{\e_n}\G^{a}_{c,*}\Vert_{L^{\infty}((\e_{n}\math{Z})^{d})}
+
\Vert \dpr^{\a}_{\e_n} \G^{a}_{c,*}-
\dpr^{\a}_{c}\G^{a}_{c,*}\Vert_{L^{\infty}(\math{R}^{d})}
\Eq(1.241) 
$$ 
By Sobolev embedding followed by Theorem~2.1 with $l=n$ and $k$
sufficiently large, the first term is bounded as required by the right
hand side of \equ(1.24) so we now consider the second term.  We bound
the $L^{\infty}$ norm by the $L^{1}$ norm of the Fourier transform.
The derivatives $\dpr^{\a}_{\e_n} - \dpr^{\a}_{c}$ give rise to a
factor
$$
\bigg|
\prod_{j=1}^{d}
\Big(
\epsilon_{n}^{-1}\big(e^{i\epsilon_{n} k\cdot e_{j}}-1\big)
\Big)^{\a_{j}}
-
\prod_{j=1}^{d}
\big(i k\cdot e_{j} \big)^{\a_{j}}
\bigg|
\le
{|\a|\over 2} \epsilon_{n}|k|^{|\a|+1}
$$
in the $L^{1}$ norm of the Fourier transform. The desired result then
follows from the continuum version of \equ(1.14).\bull 

\\Applying the above to the sequence of L\'evy fluctuation covariances  
$\{\G_{n}\}_{n\ge 0}$ we have

\vglue0.2cm
\\{\it Corollary~2.3}: Under the same conditions as above we have for all 
$k\ge 0$

$$\Vert \G_{n}-\G_{c,*}\Vert_{L^{1}_{k}((\e_{l}\math{Z})^{d})}
\le c_{k,L}L^{-{n\over 2}}.\Eq(1.25)$$

\\Moreover for all $|\a|\ge 0$

$$\Vert \dpr^{\a}_{\e_n} \G_{n}-
\dpr^{\a}_{c}\G_{c,*}\Vert_{L^{\infty}((\e_{l}\math{Z})^{d})}
\le c_{k,L}L^{-{n\over 2}}. \Eq(1.26)$$

\vglue.5truecm
\\{\bf 3. Proof of Theorem~2.1  } 
\numsec=3\numfor=1 
\vglue.5truecm

\\Theorem~2.1 follows by combining the following two lemmas whose
proofs are given below. The first is about the convergence of
continuum covariances and the second is about lattice covariances.

\vglue0.3cm

\\{\it Lemma~3.1}: For $k\ge 0$, there is a constant $c_{k,L}$ such
that for $a\ge 0$ and $n\ge 2$

$$|\hat\G^{a}_{c,n}(p)- \hat\G^{a}_{c,*}(p) |
\le c_{k,L}\>e^{-ca^{1\over 2}} (1+p^2)^{-k} L^{-{n\over 2}}. \Eq(2.1)$$

\vglue0.3cm

\\{\it Lemma~3.2}: Let $p\in B_{\e_n}$.  Then for all $n\ge d$ and all
$k\ge 0$, $a\ge 0$ there is a constant $c_{k,L}$ independent of $n$
such that

$$|\hat\G^{a}_{n}(p)- \hat\G^{a}_{c,n}(p) |
\le c_{k,L}\>e^{-c a^{1\over 2}}  (1+p^2)^{-k}(1+(bp^2 +a)^{-1})L^{-n} \Eq(2.2)$$

\\where $b$ is a positive constant independent of $n$ and other parameters.

\\{\it Proof of Theorem~2.1}: Since Lemmas~2.1 and 2.2 hold for all $k\ge 0$
we can replace $k$ by $k+d+1$. From Lemma~3.1 and Lemma~3.2 we get

$$|\hat\G^{a}_{n}(p)- \hat\G^{a}_{c,*}(p) |
\le c_{k,L}e^{-c a^{1\over 2}}  (1+p^2)^{-(k+d+1)}(1+(bp^2 +a)^{-1})L^{-{n\over 2}}
\Eq(2.222)$$

\\for all $k\ge 0$. By definition 

$$\Vert \G^{a}_{n}-\G^{a}_{c,*}\Vert_{L^{1}_{2k}((\e_{l}\math{Z})^{d})}
=\Vert \G^{a}_{n}-\G^{a}_{c,*}\Vert_{L^{1}_{2k}(U_{\e_l}(6L))}=
\sum_{|\a|=2k} \int_{U_{\e_l}(6L)} dx |D^{\a} (\G^{a}_n(x)
-\G^{a}_{c,*}) |. $$

\\After introducing a Fourier transform we get 

$$\Vert\G^{a}_{n}-\G^{a}_{c,*}\Vert_{L^{1}_{2k}((\e_{l}\math{Z})^{d})}
\le \sum_{|\a|=2k} \int_{U_{\e_l}(6L)} dx \int_{B_{\e_l}} 
{dp\over (2\p)^{d}} |D^{\a} e^{ipx}|\>
|\hat\G^{a}(p)-\hat\G^{a}_{c,*}(p)|.$$

\\Using the definition of lattice derivatives and multiderivatives we
get the trivial inequality 

$$|D^{\a} e^{ipx}|\le \prod_{j=1}^{d} |p_j|^{\a_j} \le
(p^{2})^{{|\a|\over 2}} $$

\\which we use to majorize the inequality preceding it by 

$$\Vert\G^{a}_{n}-\G^{a}_{c,*}\Vert_{L^{1}_{2k}((\e_{l}\math{Z})^{d})}
\le c_{k,L} \int_{B_{\e_l}} {dp\over (2\p)^{d}} (p^{2})^{k}  
|\hat\G^{a}(p)-\hat\G^{a}_{c,*}(p)| \Eq (2.33)$$

\\where the constant $ c_{k,L}$ depends on $L$ through the volume of
the cube  $U_{\e_l}(6L)$ and on $k$ because

$$\sum_{|\a|=2k}\>1 =c_k. $$

\\We majorize the right hand side of \equ(2.33) using the bound
\equ(2.222). Note that $d\ge 3$ so that integrability is assured
 uniformly in $l$ for all $a\ge 0$. We therefore get the bound

$$\Vert \G^{a}_{n}-\G^{a}_{c,*}\Vert_{L^{1}_{2k}((\e_{l}\math{Z})^{d})}
\le c_{k,L} L^{-{n\over 2}}e^{-ca^{1\over 2}} $$

\\which proves Theorem~2.1.  \bull

\vglue0.3cm

\\{\it Proof of Lemma~3.1}: We will divide the proof into two cases. 

\vglue0.3cm

\\Case 1. Suppose  $|p| > L^{n/2}$ or $a > L^{n}$. 

\\Recall that 
$\hat\G^{a}_{c,n}(p)$ and its pointwise limit
$\hat\G^{a}_{c,*}(p)$ satisfy the uniform bound \equ(1.20). Therefore

$$|\hat\G^{a}_{c,n}(p)- \hat\G^{a}_{c,*}(p) |\le
|\hat\G^{a}_{c,n}(p)|+|\hat\G^{a}_{c,*}(p) |
\le c_{k,L}e^{-ca^{1\over 2}}(1+p^2)^{-2k}. \Eq(2.3)  $$ 

\\Suppose $|p| > L^{n/2}$. Then from the above

$$|\hat\G^{a}_{c,n}(p)- \hat\G^{a}_{c,*}(p) |
\le c_{k,L}L^{-n} e^{-ca^{1\over 2}}(1+p^2)^{-k} \Eq(2.4)  $$ 

\\which gives the desired bound. 

\\Suppose now that  $a > L^{n}$. Then for
any $O(1)$ constant $c>0$ 

$$ e^{-ca^{1\over 2}} \le  e^{-{c\over 2} a^{1\over 2}}  
e^{-{c\over 2}L^{n\over 2}}\le {4\over c^{2}} L^{-n}e^{-{c\over
2}a^{1\over 2}}.$$

\\Inserting this bound in \equ(2.3) we get with a new constant
$c_{k,L}$ and a new constant $c$

$$|\hat\G^{a}_{c,n}(p)- \hat\G^{a}_{c,*}(p) |\le
c_{k,L}L^{-n} e^{-ca^{1\over 2}}(1+p^2)^{-k} \Eq(2.5)  $$ 

\\as desired.

\vglue0.3cm

\\Case 2. This is the converse of Case 1, namely $|p| \le L^{n/2}$ 
and $a \le L^{n}$.

\\From \equ(1.21) and \equ(1.18) we have 

$$\hat\G^{a}_{c,*}(p)=
=\prod_{m=n+1}^{\infty}|\hat A^{a}_{c,m}(R_m)(p)|^{2}\>\hat\G^{a}_{c,n}(p). $$

\\Therefore 

$$\eqalign{|\hat\G^{a}_{c,*}(p)-\hat\G^{a}_{c,n}(p)| &\le |\hat\G^{a}_{c,n}(p)|
\Big|\prod_{m=n+1}^{\infty}|\hat A^{a}_{c,m}(R_m)(p)|^{2} -1\Big|\cr
&\le c_{k,L}e^{-ca^{1\over 2}}(1+p^2)^{-2k}\Big|\prod_{m=n+1}^{\infty}
|\hat A^{a}_{c,m}(R_m)(p)|^{2} -1\Big|} \Eq(2.6) $$

\\where we have used the bound \equ(1.19).
From the continuum version of the estimate (6.17) on page 442 of [BGM]
we have 

$$|1-\hat A_{c,m}(p)| \le R_{m}|p| +a R_{m}^{2}$$

\\where $R_m = L^{-(m-1)}$. In the present Case 2 we have $|p| \le L^{n/2}$, 
$a \le L^{n}$, and in \equ(2.6) $m\ge n+1$ with $n\ge 2$. 
It is then easy to see that 
$|1-\hat A_{c,m}(p)|\le 2 L^{-{(m-1)\over 2}}$. Whence

$$(1-2L^{-{(m-1)\over 2}})^{2}\le |\hat A_{c,m}(p)|^{2}\le 
(1+2L^{-{(m-1)\over 2}})^{2}. \Eq(2.7)$$

\\1. From $1+x\le e^{x}$, for $x\ge 0$, and \equ(2.7) we have 
$|\hat A_{c,m}(p)|^{2}\le e^{4 L^{-{(m-1)\over 2}}}$. Therefore

$$\eqalign{\prod_{m=n+1}^{\infty}|\hat A^{a}_{c,m}(R_m)(p)|^{2} &\le
e^{4 \sum_{m=n+1}^{\infty}L^{-{(m-1)\over 2}}}
\le e^{O(1)L^{-{n\over 2}}}\cr
&\le 1+O(1)L^{-{n\over 2}}} \Eq(2.8)  $$

\vglue0.2cm \\2. It is easy to see from the lower bound in \equ(2.7)
that, for $m\ge n+1$, and $ n\ge 2$ we get

$$|\hat A^{a}_{c,m}(R_m)(p)|^{2}\ge 
e^{2\log(1-2L^{-{(m-1)\over 2}})}=
e^{-O(1)L^{-{(m-1)\over 2}}}$$

\\whence

$$\eqalign{\prod_{m=n+1}^{\infty}|\hat A^{a}_{c,m}(R_m)(p)|^{2} &\ge
e^{-O(1)\sum_{m=n+1}^{\infty} L^{-{(m-1)\over 2}}}
=e^{-O(1)L^{-{n\over 2}}(1-L^{-1/2})^{-1}} \cr
&\ge e^{-O(1)L^{-{n\over 2}}} }. $$

\\For $x\ge 0$ and sufficiently small we have $e^{-x}\ge
1-2x$. Therefore we get from the previous inequality

$$\prod_{m=n+1}^{\infty}|\hat A^{a}_{c,m}(R_m)(p)|^{2} \ge
1- O(1)L^{-{n\over 2}}.\Eq(2.9) $$

\\From \equ(2.8) and \equ(2.9) we get

$$\Big|\prod_{m=n+1}^{\infty}|\hat A^{a}_{c,m}(R_m)(p)|^{2} -1 \Big| 
\le O(1)L^{-{n\over 2}}.\Eq(2.10) $$

\\Inserting the bound \equ(2.10) in \equ(2.6) gives

$$|\hat\G^{a}_{c,*}(p)-\hat\G^{a}_{c,n}(p)|\le 
c_{k,L}L^{-{n\over 2}}e^{-ca^{1\over 2}} (1+p^2)^{-2k} $$

\\which completes the proof of Lemma~3.1.  \bull

\vglue0.3cm

\\{\it Proof of Lemma~3.2}: 

\vglue0.2cm

\\The proof of Lemma~3.2 reposes crucially on 
Lemma~6.7, [BGM, page 441],
and {\it Claim~2.3} to follow. According to Lemma~6.7 of [BGM],
for $0\le m \le n$, there is a constant $c_{L,m}$ independent of $n$
such that 

$$|\hat A^{a}_{\e_n ,m}(R_m)(p)- \hat A^{a}_{c,m}(R_m)(p)| 
\le c_{L,m} \e_n.  \Eq(2.11)  $$

\\It will be important to have a control on the $m$-dependence of the constant
$c_{L,m}$ in \equ(2.11). This is provided by 

\vglue0.2cm

\\{\it Claim~3.3}:

$$c_{L,m}= O(1) L^{-{(d-2)\over 2}m} L^{{d\over 2}}. \Eq(2.12)$$

\\Sketch of proof: Claim~2.3 follows from an 
examination of proof of Lemma~6.7, [BGM]. This proof needs the Poisson kernel
estimate (Proposition~5.2) and Lemma~6.5 both of which are proved in 
Appendix~A of [BGM]. The Poisson kernel estimate gives a constant 
$O(1)R_{m}^{d/2}$ where $R_m =L^{-(m-1)}$. An additional $L^{m}$ arises from a 
derivative on $g_{m}$  (see the proof of Lemma~6.7). Therefore
$R_{m}^{d/2}L^{m}$ is the constant of Lemma~6.7 and 
gives the right hand side of \equ(2.12). \bull

\vglue0.2cm
\\From \equ(1.16) and \equ(1.18)

$$\hat\G^{a}_{n}(p)- \hat\G^{a}_{c,n}(p)=
|\hat\AA^{a}_{n}(p)|^{2}\Big(\hat\G^{a}_{\e_n}(p)-\hat\G^{a}_{c}(p)\Big)+
\Big(|\hat\AA^{a}_{n}(p)|^{2}-|\hat\AA^{a}_{c,n}(p)|^{2}\Big)\hat\G^{a}_{c}(p)
$$
 
\\whence on using the bounds \equ(1.15), \equ(1.20) together with (see
[BGM, page 435 ])

$$|\hat\G^{a}_{c}(p)|\le c_{L}(1+p^{2})^{-1} $$

\\and $|A^{2}-B^{2}| \le |A-B||A+B|$, we get

$$|\hat\G^{a}_{n}(p)- \hat\G^{a}_{c,n}(p)|\le 
c_{k,L}e^{-O(1)a^{1\over 2}} (1+p^2)^{-k}
\Big(|\hat\G^{a}_{\e_n}(p)-\hat\G^{a}_{c}(p)| +
|\hat\AA^{a}_{n}(p)-\hat\AA^{a}_{c,n}(p)|\Bigr). \Eq(2.13) $$

\\We will estimate the two terms within the big round brackets above.

\vglue0.2cm
\\1. From \equ(1.9) and the continuum analogue of \equ(1.9) we get

$$\hat\AA^{a}_{n}(p)-\hat\AA^{a}_{c,n}(p) = $$
$$=
\sum_{m = 1}^{n}\quad 
\prod_{i<m} \hat A^{a}_{\e_n ,i}(R_i)(p)
\Big(\hat A^{a}_{\e_n ,m}(R_m)(p)- \hat A^{a}_{c,m}(R_m)(p)\Big)
\prod_{j>m}\hat A^{a}_{c,j}(R_j)(p). $$

\\We bound $|\hat A^{a}_{\e_n ,i}(R_i)(p)|$ and $|\hat
A^{a}_{c,j}(R_j)(p)|$ by $1$ and the $m$ factor by \equ(2.11) and
\equ(2.12). We get

$$|\hat\AA^{a}_{n}(p)-\hat\AA^{a}_{c,n}(p)|\le
\sum_{m = 1}^{n}
\quad L^{-(n-{d\over 2})}  
O(1)L^{-{(d-2)\over 2} m}. 
$$

\\By the conditions of Lemma~3.2, we may take $L$ sufficiently large
and $d\ge 3$ so that the series is geometrically convergent and
dominated by the first term.  Therefore

$$|\hat\AA^{a}_{n}(p)-\hat\AA^{a}_{c,n}(p)|\le O (L) L^{-n}.\Eq(2.14)$$

\vglue0.2cm
\\2. We now estimate the first term within the big round brackets in 
\equ(2.13). From \equ(1.7) and \equ(1.17) we get

$$\eqalign{\hat\G^{a}_{\e_n}(p)-\hat\G^{a}_{c}(p)=&
(\hat G^{a}_{\e_n}(p)-\hat G^{a}_{c}(p))\>
(1-|\hat A^{a}_{c,0}(L)(p)|^{2})+ \cr
&\hat G^{a}_{\e_n}(p)\>
\big(|\hat A^{a}_{c,0}(L)(p)|^{2}- |\hat A^{a}_{\e_{n},0}(L)(p)|^{2}\big).} $$

\\Using the bounds $|\hat A^{a}_{c,0}(L)(p)|\le 1$ and 
$|\hat A^{a}_{\e_{n},0}(L)(p)|\le 1$ we get

$$|\hat\G^{a}_{\e_n}(p)-\G^{a}_{c}(p)| \le 
2\>|\hat G^{a}_{\e_n}(p)-\hat G^{a}_{c}(p)|
+  2\>|\hat G^{a}_{\e_n}(p)|\> |\hat A^{a}_{c,0}(L)(p)
- \hat A^{a}_{\e_{n},0}(L)(p)|. \Eq(2.15) $$

\\We first bound the second term in \equ(2.15). There exists a constant
$b$ independent of $n$ such that (see equation (5.9) in  [BGM], page
434, we have replaced the constant $c$ by $b$)

$$0\le \hat G^{a}_{\e_n}(p)\le (a +bp^{2})^{-1}. \Eq(2.16) $$

\\Furthermore from \equ(2.11)

$$|\hat A^{a}_{\e_n ,0}(L)(p)- \hat A^{a}_{c,0}(L)(p)| 
\le c_{L} \e_n.  \Eq(2.17)  $$

\\Therefore

$$|\hat G^{a}_{\e_n}(p)|\> |\hat A^{a}_{c,0}(L)(p)
- \hat A^{a}_{\e_{n},0}(L)(p)|\le c_{L}(a+bp^{2})^{-1}\e_n. \Eq(2.18)$$ 

\\Next we bound the first term on the right hand side of \equ(2.15).

\vglue0.2cm

\\1. Consider first the case $|p|\ge L^{n/2}$. Then from \equ(2.16) and 
$\hat G^{a}_{c}(p)=(a+p^{2})^{-1}$ we we get

$$|\hat G^{a}_{\e_n}(p)-\hat G^{a}_{c}(p)|\le |\hat G^{a}_{\e_n}(p)| +
|\hat G^{a}_{c}(p)|\le O(1)L^{-n}. \Eq(2.19)  $$

\\2. Next we consider the case $|p| <  L^{n/2}$. From the definition above
of  $\hat G^{a}_{c}(p)$, and $\hat G^{a}_{\e_n}(p)$ we get

$$ \hat G^{a}_{\e_n}(p)-\hat G^{a}_{c}(p)=
{p^{2} +\hat\D_{\e_n}(p)\over (a+p^{2})(a-\hat\D_{\e_n}(p))}. $$

\\Now using the bound \equ(2.16) we get

$$|\hat G^{a}_{\e_n}(p)-\hat G^{a}_{c}(p)| \le 
O(1){1\over (p^{2})^{2}}|p^{2} +\hat\D_{\e_n}(p)|. $$

\\From $|p| <  L^{n/2}$ and $\e_n=L^{-n}$ we have $\e_n|p|<L^{-n/2}$.
Now expanding out $\hat\D_{\e_n}(p)$ (see \equ(1.2)) in an absolutely
convergent series we easily get the estimate for \Red{***}
$n\ge d$

$$|p^{2} +\hat\D_{\e_n}(p)|\le O(1)\e_{n}^{2}(p^{2})^{2}. $$

\\Combining this with the earlier inequality we get 

$$|\hat G^{a}_{\e_n}(p)-\hat G^{a}_{c}(p)| \le O(1)\e_{n}^{2}. \Eq(2.20) $$

\\From \equ(2.19) and \equ(2.20) we get for all $p\in B_{\e_n}$

$$|\hat G^{a}_{\e_n}(p)-\hat G^{a}_{c}(p)| \le O(1)\e_{n}. \Eq(2.21) $$

\\Inserting the bounds \equ(2.18) and \equ(2.21) in \equ(2.15) we get

$$ |\hat\G^{a}_{\e_n}(p)-\G^{a}_{c}(p)| \le
c_{L}\e_{n}(1 +(a+bp^{2})^{-1}). \Eq(2.22) $$

\\From \equ(2.13) and the bounds \equ(2.14) and \equ(2.22) we get

$$|\hat\G^{a}_{n}(p)- \hat\G^{a}_{c,n}(p)|\le 
c_{k,L}
e^{-ca^{1\over 2}}
L^{-n} (1+p^2)^{-k}
(1 +(a+bp^{2})^{-1}). \Eq(2.22)   $$

\\This proves Lemma~3.2. \bull

\vglue.5truecm
\\{\bf {Acknowledgement:}} 
The work of DB was supported in part by NSERC of
Canada. DB thanks the Institute for Advanced Study for membership while this work was in progress.

\vglue.5cm

\\{\bf References}

\vglue.3cm
\\[A] R.A. Adams, Sobolev Spaces, Academic Press, Inc. (London)
 1975

\vglue.3truecm  
\\[BGM] D. Brydges, G. Guadagni and P. K. Mitter: Finite range
Decomposition of Gaussian Processes,
J Stat Phys (2004) {\bf 115}: 415--449  

\vglue.3truecm
\\[BS] D. Brydges, G. Slade: Renormalization goup analysis of weakly
self-avoiding walk in dimension four and higher,
Proceedings of the International Congress of Mathematicians (2010),
Hyderabad, India, http://arxiv.org/pdf/1003.4484

\vglue.3truecm
\\[BT] D. Brydges and A. Talarczyk: Finite Range Decomposition of
Positive Definite Functions,
J Funct Anal (2006)  {\bf 236}: 682-711 

\vglue.3cm
\\[D] J. Dimock: Infinite volume limit for the Dipole Gas,
J Stat Phys (2009) {\bf 135}: 393-427 

\vglue.3cm
\\[F] P. Falco: Kosterlitz-Thouless Transition Line for the Two
Dimensional Coulomb Gas, http://arxiv.org/abs/1104.1974

\vglue.3truecm
\\[MS] P.K. Mitter and B. Scoppola: The Global Renormalization Group
Trajectory in a Critical Supersymmetric Field Theory on the Lattice $\math{Z}^{3}$, 
J. Stat Phys (2008) {\bf 133}: 921-101

\end